\documentclass[aps,prb,twocolumn,showpacs,floatfix]{revtex4}
\usepackage{graphicx}

\begin{document}

\title{Ground state phase diagram of a spinless, extended Falicov-Kimball model
on the triangular lattice}
\author{Umesh K. Yadav, T. Maitra, Ishwar Singh}
\affiliation{Department of Physics, Indian Institute of Technology Roorkee, Roorkee- 247667, Uttarakhand, India}
\author{A. Taraphder}
\affiliation{Department of Physics and Centre for Theoretical Studies, Indian Institute of Technology, Kharagpur - 721 302, India}
\date{\today}

\begin{abstract}
Correlated systems with hexagonal layered structures have
come to fore with renewed interest in Cobaltates, transition-metal
dichalcogenides and $GdI_2$. While superconductivity, unusual metal
and possible exotic states (prevented from long range order by strong 
local fluctuations) appear to come from frustration and correlation working 
in tandem in such systems, they freeze at lower temperature to 
{\it crystalline} 
states. The underlying effective Hamiltonian in some of these systems 
is believed 
to be the Falicov-Kimball model and therefore, a thorough study of the ground 
state of this model and its extended version on a non-bipartite lattice 
is important. Using a Monte Carlo search algorithm, we identify a large 
number of different possible ground states with charge order as well as valence 
and metal-insulator transitions. Such competing states, close in energy, 
give rise to the complex charge order and other broken symmetry structures as 
well as phase segregations observed in the ground state of these systems. 

\end{abstract}
\vspace{0.5cm} 
\maketitle
\section{Introduction}

Systems like transition-metal dichalcogenides~\cite{aebi,qian2,cava}, 
cobaltates~\cite{qian},
$GdI_{2}$~\cite{gd1,gd2} and its doped variant $GdI_{2}H_{x}$~\cite{gd3} have 
attracted considerable attention recently as they exhibit
a number of remarkable cooperative phenomena, like valence and metal insulator 
transition, charge and magnetic order, excitonic instability~\cite{aebi} and 
possible non-Fermi liquid states~\cite{gd2,castro}.  
The underlying geometric frustration of the triangular lattice, coupled
with strong quantum fluctuations, gives rise to a large degeneracy 
at low temperatures and competing ground states close by in energy.   
A consequence of this is a fairly complex ground state magnetic phase diagram
~\cite{gd3} and the presence of soft local modes strongly coupled to
the conduction electrons~\cite{gd2}.

These systems are characterized by the presence of localized and  
itinerant electrons confined to the two-dimensional triangular lattice.
The electronic coherence along the perpendicular direction is negligible owing to 
the presence of large intervening ions (like Iodine in $GdI_2$) between the 
layers of rare earth or transition metal ions. It has been suggested recently 
that these correlated systems may very well be described by 
an effective Falicov-Kimball model (FKM) on a triangular lattice. The 
nearly degenerate manifold of $d$-states close to the Fermi level, seen in
LSDA calculations~\cite{gd3}, breaks down into a narrow, core-like state
pushed below the Fermi level and an extended state crossing it, due to  
the strong local correlations in the $d$-manifold~\cite{gd2}. The situation is 
similar to the well-known FKM~\cite{fkm}, if electrons in this narrow, localized band
are assumed to posses a very large effective mass. In the usual FKM one considers 
only the kinetic energy of the itinerant electrons and the local Coulomb 
interaction between the itinerant and localized electrons. This model has
a long history and has been used to study metal insulator
transition, magnetism in correlated systems and in the version where the
interaction is attractive (the sign of the interaction is irrelevant in a 
particle-hole symmetric case), it can be used to model crystallization 
in solids~\cite{lieb}. Most of the studies are mean-field in nature. There
are only a few exact results which are almost entirely on bipartite lattices
~\cite{brandt,free,lach,kennedy2} and are in lower dimensions~\cite{free2}. 

A very important question, in this context, is the nature of the ground states
in such a system. Most of the systems described above show charge and orbital
order at low temperatures. The layered material Na$_x$CoO$_2$, which shows
superconductivity on hydration, reveals a $\sqrt {3}\times \sqrt{3}$ 
superlattice~\cite{qian} at low temperatures due to spacial modulation of 
charge/orbital densities. Transition metal dichalcogenides mostly freeze into 
charge density wave states~\cite{aebi_jes}, both commensurate and incommensurate 
(on doping some of them become superconductors at low temperature).      
Superlattice spots have also been seen in $GdI_2$ in X-ray scattering at 
low temperatures. Modelling the low-energy physics of some of these materials 
in an effective FKM naturally leads to such ordered states at low temperature.
The FKM is known to support various kinds of ordered or 
phase segregated~\cite{free,gruber} ground states on a bipartite lattice. 
Some of these ordered structures appear in the ground state of a triangular 
lattice as well~\cite{gruber}. There are hardly any rigorous results for the 
ground state of FKM on a triangular lattice. Most of 
the results are perturbative~\cite{gruber,wojt}, valid for large $U$. The 
combined effects of correlation and geometric frustration  
on the ground state phase diagram have not been investigated extensively using 
numerical methods on finite size lattices either. With the appearance of 
charge ordered states as the low-energy broken symmetry ground states
in the systems described above, it is imperative to look for the solutions
of the FKM at low temperatures. In this work, therefore, we study the FKM in 
its spinless version numerically for all ranges of interaction, and explore 
the ground state charge ordering on a triangular lattice.  

It may be noted that in some rare earth compounds (specially the mixed-valence 
compounds), the rare earth ions with two different
ionic configurations $f^{n}$ and $f^{n-1}$ have different ionic radii. The 
contraction of $d$-orbitals in ions having $f^{n-1}$ configuration, for reduced 
screening of nuclear charge, leads to increased localization of them.  
This implies that the $d$-orbital overlap between nearest neighbors depend on 
local $f$-electron configurations of neighboring ions, resulting in a correlated 
hopping of $d$-electrons. Such a correlated hopping term also appears in the first 
principles calculation~\cite{hirsch} of the tight binding 
Hamiltonian and is usually neglected in Hubbard type models for being smaller 
compared to the on-site term. Its significance in superconductivity has been 
emphasized already~\cite{hirsch2,bulk} and in the context of FKM, it has been 
considered by several authors~\cite{wojt,shv,farkov} using various approximations.  
We, therefore, extend the model to include a correlated hopping term in 
the Hamiltonian for the itinerant electrons.  

\begin{eqnarray}
{H} =-\sum_{\langle ij\rangle}(t_{ij}+\mu\delta_{ij})d^{\dagger}_{i}d_{j}+E_f \sum_{i} f^{\dagger}_{i}f_{i}
\nonumber \\
+U \sum_{i}f^{\dagger}_{i}f_{i}d^{\dagger}_{i}d_{i}+\sum_{\langle ij\rangle}t^{\prime}_{ij}(f^{\dagger}_{i}f_{i}+f^{\dagger}_{j}f_{j})
d^{\dagger}_{i}d_{j}.
\end{eqnarray}

Here $d^{\dagger}_{i}, d_{i}\,(f^{\dagger}_{i}, f_{i}$)  are, respectively, 
the creation
and annihilation operators for itinerant $d$-electrons (localized $f$-electrons) at 
the site $i$. The first term in Eq.(1) is the kinetic energy of $d$-electrons 
on a triagular lattice: only nearest-neighbor hopping is considered. The second 
term represents the dispersionless energy level $E_{f}$ of the $f$-electrons while 
the third term is the on-site Coulomb repulsion between $d$- and $f$-electrons. The 
last term, a correlated hopping term, is an extension over the usual FKM as 
discussed above. In the context of $GdI_2$ (or the transition metal dichalcogenides),
the $f$-electrons represent the localized (or valence) band below the Fermi
level while $d$-electrons cross the Fermi level in the effective low-energy model
alluded to earlier~\cite{gd2,at_tobe}.

\section{Methodology}

The Falicov-Kimbll Hamiltonian (1), conserves the local $f$-electron occupation
number $\hat{n}_{f,i}=f^{\dagger}_{i}f_{i}$ owing to its local $U(1)$ gauge
invariance in the absence of $f-d$ hybridization. Therefore, $[\hat{n}_{f,i},H]=0$
and $\omega_{i}=f^{+}_{i}f_{i}$ is a good quantum number taking values only 0 or 1.
The gauge symmetry also implies~\cite{elitzur} interband excitonic averages of the 
type $<d_{i}^{\dagger}f_{j}>$ are identically zero at any finite temperature 
(implying an absence 
of homogeneous mixed valence) and the $f$-electron level remains dispersionless 
($\langle f_{i}^{\dagger}f_{j}\rangle =0$, \, for\, $i\ne j$). On a 
non-bipartite lattice as we consider here, there is no particle-hole symmetry 
either. The local conservation 
of $f$-electron number implies that the Hamiltonian may be written as,

\begin{equation}
H=\sum_{\langle ij\rangle}h_{ij}(\omega)d^{+}_{i}d_{j}+E_f\sum_{i}\omega_{i}
\end{equation}
\noindent where $h_{ij}(\omega)=[-t_{ij}+t^{\prime}_{ij}
(\omega_{i}+\omega_{j})]+(U\omega_{i}-\mu)
\delta_{ij}$.

We set the scale of energy with $t_{<ij>}=1$. The Hamiltonian Eq.(2) represents
a non-interacting $d$-electron moving in an annealed disordered background of 
the $f$-electrons~\cite{kennedy}. The interactions among the electrons is, 
therefore, kinematical in nature, coming entirely from the Fermi statistics 
for a given $f$-electron configuration. It suffices therefore to obtain
the spectrum of this Hamiltonian for different configurations $\omega=\{\omega_1, 
\omega_2, \dots , \omega_N\}$ of $f$-electrons by numerically diagonalizing the
Hamiltonian over different $\omega$ and annealing over the configurations.
We perform this on a triangular lattice of finite size with periodic boundary 
conditions (PBC). Although this entails a
tremendous simplification over solving a full interacting quantum Hamiltonian,
even for a relatively small lattice size (in this work we considered $L \times L$
lattice with $L=12$ and checked the results at larger $L$ values in test cases), 
the number of different $f$-electron configurations is exponentially large and it may not be feasible to explore all the configurations within a reasonable 
computer time. For example, even for N=$36$ and $N_{f} =18$, the number 
of configurations $N_{C_{N_{f}}} \sim 10^{10}$. 
\begin{figure*}[ht]
\begin{center}
\includegraphics[width=14cm]{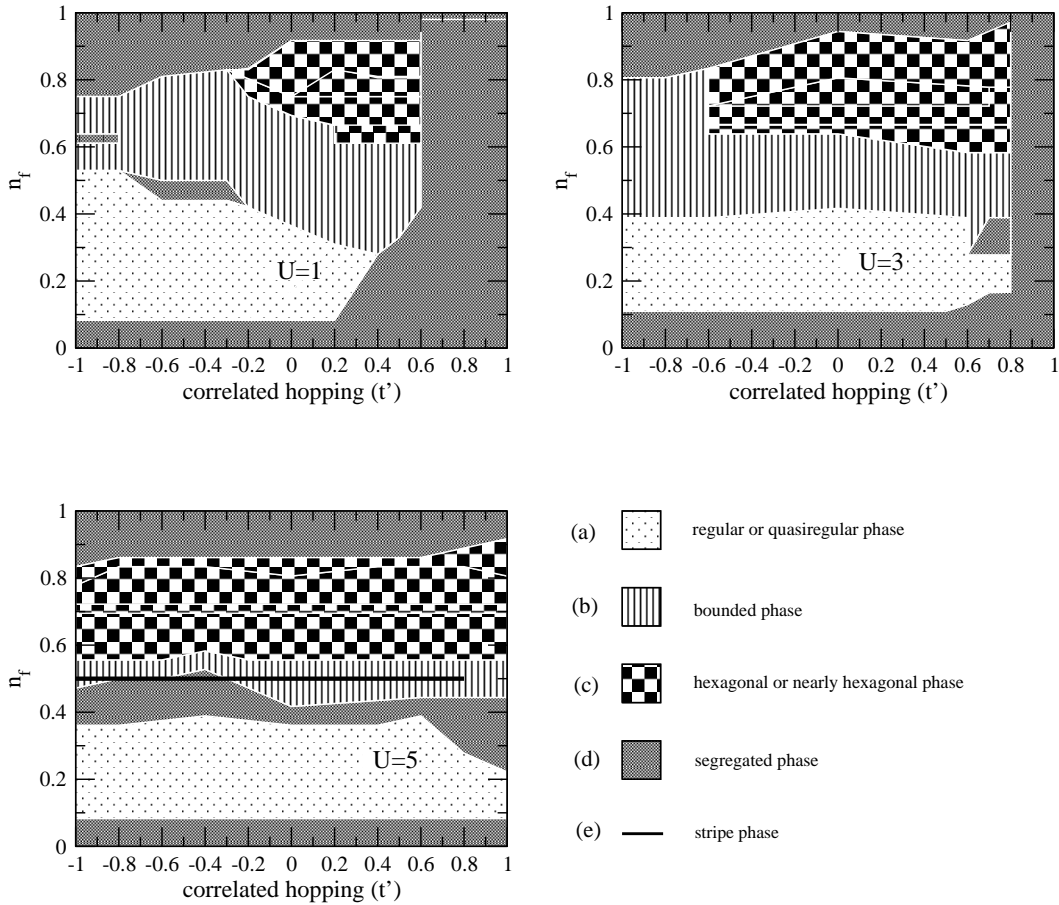}
\caption{$n_{f}-t^{\prime}$ phase diagram for $U=1$, $U=3$ and $U=5$. In 
the $U=5$ phase diagram there is a {\it stripe} phase exactly 
at $n_f=1/2$ from $t^{\prime}=-1$ to $0.8$ (shown by a horizontal line from  
$t^{\prime}=-1$ to $t^{\prime}=0.8$) } \label{gif}
\end{center}
\end{figure*}

Therefore, one needs efficient search algorithms and explore only a relatively 
small subset of the entire $f$-electron configurations. We have used a Monte Carlo
sampling method to achieve this goal. We work at half-filling, 
i.e., $N_{f}+N_{d}=N$ where $N_{f},\, N_{d}$ are the total number of $f$- 
and $d$-electrons and $N=L^2$ is the number of sites. 
For a lattice of $N$ sites, the basis is chosen as $d_{1}^\dagger\,|0>,\cdots, 
d_{N}^{\dagger}\,|0>$ and $H$ (in Eq.(2)) is now an $N\times N$ matrix for a
fixed configuration $w$. The partition function is, therefore  
 
\begin{equation}
{\cal Z}=\prod_{i}\left(\sum_{{\omega_i=0,1}}\, 
Tr\, e^{ -\beta H(\{\omega_i\})} \right ) 
\end{equation}
\noindent where the trace is taken over the $d$-electrons, and $\beta=\frac{1}{kT}$.
The trace can be calculated from the eigenvalues $\lambda_{i}\, (i=1\cdots N)$ of 
the matrix $h$ (first term in $H$ (Eq(2)) quite easily. $h$ is diagonalized by a 
unitary transformation $U^{\dagger}hU=K$, where $K$ is 
a diagonal matrix with $\lambda_i$ its diagonal elements. The unitary rotation 
gives the diagonal basis (in which $h$ is diagonal) represented
by the eigenvectors $v_{1}^\dagger\,|0>,\cdots, v_{N}^{\dagger}\,|0>$. Defining
the operator $\hat{n_i}=v_{i}^{\dagger}v_{i}$ and writing $n_i$ to denote the
corresponding eigenvalues of $\hat{n_i}$, the trace above can be identified 
with 
 
\begin{eqnarray}
Tr\, e^{ -\beta h} &=&
\sum_{n_{1}\cdots n_{N}} < n_{1}\cdots n_{N}|e^{-\beta h}| n_{1}\cdots n_{N}>
\nonumber \\
&=& \sum_{n_{1}\cdots n_{N}} < n_{1}\cdots n_{N}|e^{-\beta \sum_{i}\lambda_{i}n_{i}}|
n_{1}\cdots n_{N}> 
\nonumber \\
\end{eqnarray}

\noindent This reduction of the exponential to a c-number is the essential 
simplification that makes the problem amenable to a {\it classical} MC calculation. 
The partition function can, therefore, be recast in the form  
 
\begin{eqnarray}
{\cal Z}=\prod_{i}(\sum_{\omega_i=0,1}e^{-\beta E_{f}\omega_{i}})\,  
\prod_{j=1}^{N}\,(Tr_{j}e^{-\beta \lambda_{j} n_{j}})
\nonumber \\ 
=\prod_{i}(\sum_{\omega_i=0,1}e^{-\beta E_{f}\omega_{i}})\,\prod_{j}^{N}
(1+e^{-\beta\lambda_{j}})  
\end{eqnarray}

\begin{figure*}[ht]
\begin{center}
\includegraphics[width=14cm]{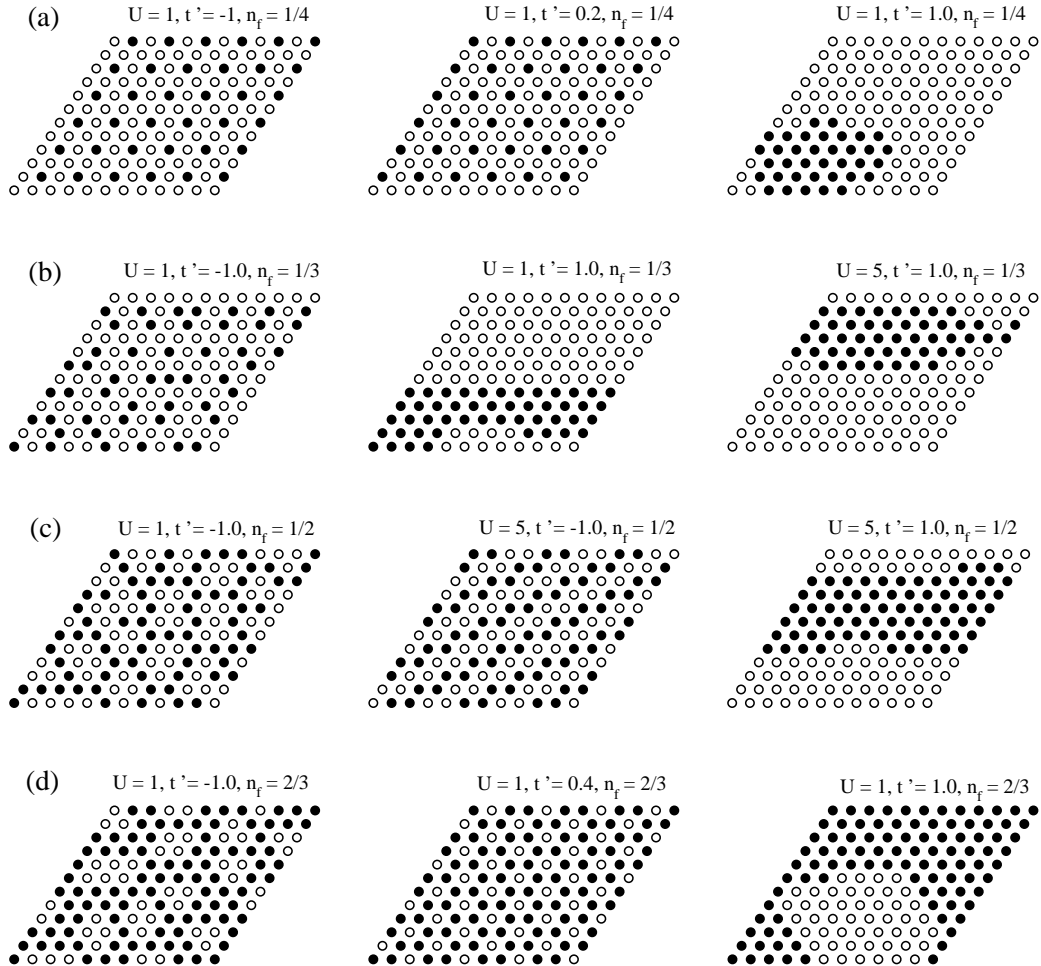}
\caption{The ground-state configurations for various values of $U$, 
$t^{\prime}$ and $n_{f}$.
Filled circles correspond to sites occupied by $f$-electrons and open circles 
correspond to unoccupied sites.}\label{gif}
\end{center}
\end{figure*}

\noindent An average of any operator operating on the $d$-electrons is then 
calculated using the usual statistical mechanical procedure $ < \hat{A}>=
Tr(\hat{A}\,e^{-\beta H})/ Tr(e^{-\beta H})$. Indeed, the intensive part
of the calculation involves finding the eigenvalues of $h$. The summations
over the static variables $\{\omega_i\}$ are evaluated using a classical Monte 
Carlo simulation. In order to get the ground state, a simulated
annealing is employed ramping the temperature down from a high to a very low 
value. The process, therefore, involves the following steps:   
(i) For a lattice of size $L\,\, (N=L^{2})$, choose a particular value
of $N_{f}$ ($0 \le N_{f} \le N$, and  $N_{d}=N-N_{f}$).
(ii) For a fixed $N$ and $N_{f}$, we choose a random configuration 
$\omega=\{\omega_1,\omega_2, \dots , \omega_N \}$ (iii) Choosing values 
for $t^{\prime}, U$ and $E_{f}$, we find the eigenvalues $\lambda_{i}$ 
of $H(\omega)$ and the corresponding total free energy $F(\omega)=-kT\, \log\, 
{\cal Z}$. The chemical potential $\mu$ is used to fix the $d$-electron 
number $N_d$. The corresponding total energy is then $E(\omega)=\lim_{T\rightarrow 0}
F(\omega)$ (iv) Generate a new random configuration $\omega^{\prime}$ and calculate 
the new energy $E(\omega^{\prime})$. 

\begin{figure}[ht]
\begin{center}
\vspace{0.5cm}
\includegraphics[width=8cm]{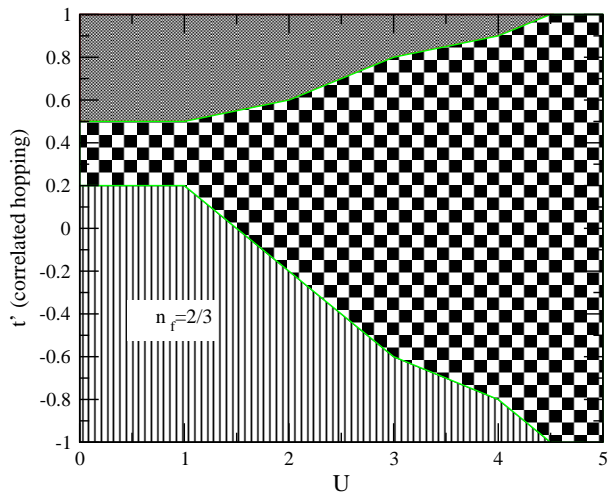}
\caption{ $U-t^{\prime}$ phase diagram at the $f$-electron 
filling $n_f=2/3$. Symbols showing the phases have the same meaning as in Fig. 1} 
\label{fig3}
\end{center}
\end{figure}
(v) Find $\Delta E=E(\omega)-
E(\omega^{\prime})$, compare $ s=e^{-\Delta E/KT}$ with a random 
number $ r_{n}$ \, ($0  <  r_{n}  <  1$):  if $  r_{n}  <  min \,(1,s)$, accept the new 
configuration $\omega^{\prime}$, else it is rejected. The steps (ii) to (v) 
are repeated until convergence is reached at a temperature $T$. The temperature is 
then ramped down slowly to a very low (compared, again, to typical $\Delta E$) 
value and at each step of $T$, the same routine is performed. Repeating this over
a few times usually lead to a unique low-energy state to the lowest temperatures
searched. Although this is still a finite temperature state, the nearest
excitation $\Delta E$ was checked to be higher than $KT$ indicating that this
is the likely ground state. This procedure converges to the `ground state' faster  
(even for the largest lattice sizes we studied) in  
contrast with other methods~\cite{farkov} and makes it possible to test 
our results in larger lattices. There are, of course, the usual problems of 
simulated annealing that one needs to take care. While ramping down in $T$, 
it often gets stuck in a local minimum even when the ramping is made ever slower 
at lower $T$. Different kinds of moves and at times simultaneous moves  
involving several sites (and at longer ranges) were employed to drive the 
system out of the quench. The procedure has been successfully used
in the search for complex ground state stuctures in manganites~\cite{dagbook}. 
It allows us to work in a much smaller sample-space of 
the whole configuration space, and any physical variable may be obtained by 
averaging over this sample space. Similar approach was employed in the study of
the phase transitions in FKM on a bipartite lattice earlier~\cite{maska,vries}.  

\section{Results and discussion}

Perturbative results for large $U$ indicate that to order $1/U$, the FKM can be
mapped onto an Ising antiferromagnet (AFM) in a magnetic field, 
$H_{eff}=\sum_{<ij>} \frac{t^2}{4U}
s_{i}s_{j}+\frac{1}{2}(\mu+E_{f})\sum_{i}s_{i}+{\rm constant  \, terms}$, where
$s_{i}=2\omega_{i}-1,\, s_{i} = -1,1$.   
The Ising AFM state on a triangular lattice is frustrated and leads to large
degeneracies at low temperature as discussed above. It turns out that this
frustration is lifted~\cite{gruber} in the higher order perturbation in $1/U$. It 
is therefore quite intriguing that one would expect the effects of frustration
to bear on the ground states as $U$ and the chemical potentials are varied. 
Different regions of the phase space are, therefore, 
controlled by different dominant effects. We studied the ground state 
configurations and the possible valence transitions as a function of a range 
of values of $U,\,$ and filling $n_{f}=N_{f}/N$ \, ($N_d$ is constrained to
$N-N_f$) using the method outlined in 
the previous section. We look at the effect of correlated hopping of $d$-electrons
on these phases and obtain the ground state configuration for several values 
of $t^{\prime} \in [-1,1]$ at a fixed $U$. 

In Fig.1 we present the phase diagram in $n_{f}$ - $t^{\prime}$ plane 
at three representative $U$ values ($U=1, 3$ and $5$).
We observe that the ground state configurations of $f$-electrons for a fixed 
filling are significantly affected by the correlated hopping $t^\prime$ and 
undergo several phase transitions at lower $U$ values. At larger $U$, the 
ground states remain nearly independent of $t^\prime$. Various ground state 
phases encountered in Fig.1 are mainly; (a) {\it Regular} or {\it quasi regular} 
phase, found around $n_{f}  = \frac{1}{4}$ (Fig.2a columns 1, 2), where 
the filled sites are arranged in an almost regular pattern. (b) {\it Bounded} 
phase, where regions of empty sites are unconnected (and always surrounded by 
filled sites), observed  around $n_f= \frac{1}{2}$ (Fig.2c and 2d, first column).
\begin{figure}[ht]
\begin{center}
\includegraphics[width=8cm]{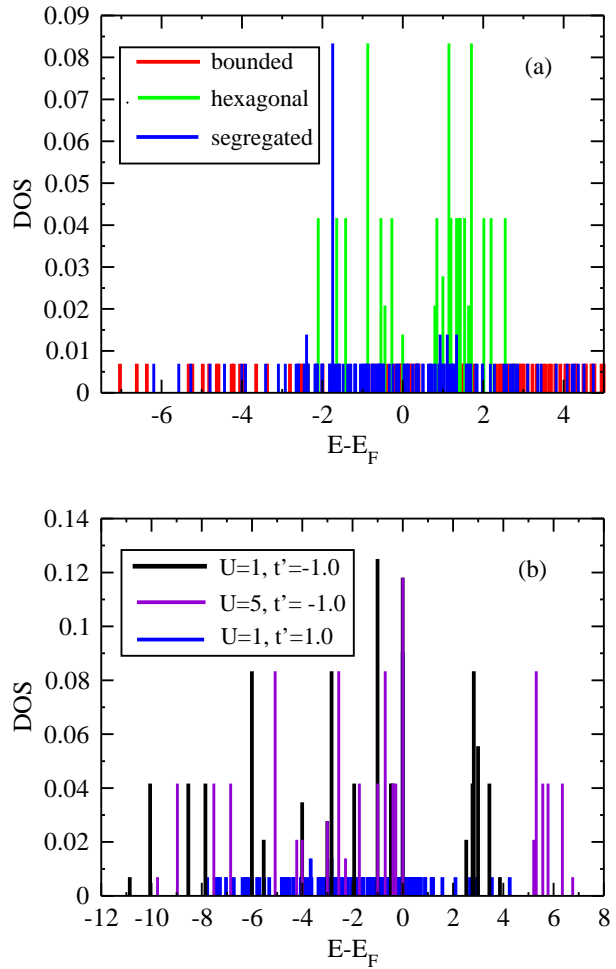}
\caption{(color online) (a) Density of states for 
three different phases at $n_f=2/3$ as $t^{\prime}$ is changed. 
Bounded: $U=1, t^{\prime}=-1$ (red), hexagonal: $U=1, t^{\prime}=0.4$ 
(green) and segregated: $U=1, t^{\prime}=1$ (blue).
(b) Density of states for two different 
phases at $n_f=1/4$. Regular: $U=1, t^{\prime}=-1$ (black) and 
$U=5,t^{\prime}=-1$ (violet)    
and segregated: $U=1,t^{\prime}=1$ (blue). E$_F$ is
the corresponding Fermi level in each case.}
\label{fig4}
\end{center}
\end{figure} 
There is no apparent spatial order in this state and it is found to shrink as $U$ 
increases. (c) {\it Hexagonal} or {\it nearly Hexagonal} phase (Fig.2d second 
column)   
where $f$-electrons form hexagonal structures, are observed at all investigated $U$ 
values. They appear at values of $n_{f}$ around $\frac{2}{3}$ and $\frac{3}{4}$.
(d) {\it Segregated phase}, where domains of sites occupied by $f$-electrons 
are segregated from the unoccupied sites (Fig.2 column 3 shows four such 
configurations). There can also be multiple such domains (not shown here). 
\begin{figure*}[ht]
\begin{center}
\includegraphics[width=14cm]{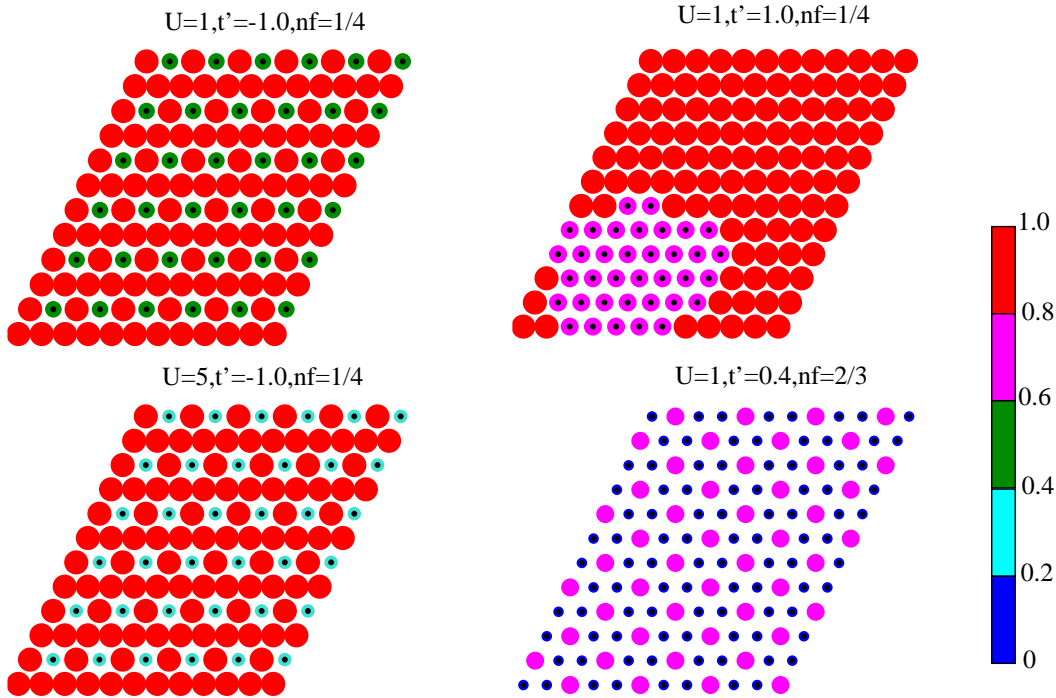}
\caption{(color online) $d$-electron densities are shown on each site 
for $U=1,\,t^{\prime}=-1.0,\, 
n_f=1/4; \,\, U=1,\, t^{\prime}=1.0,\, n_f=1/4; \,  U=5, \, t^{\prime} = -1.0, \, 
n_{f}=1/4\,$ and $U=1, \,t^{\prime}=0.4, \,n_f=2/3$. The color 
coding and the radius of the circles indicate the $d$-electron density 
profile. The dark dot indicates $\omega_{i}=1$ at that site.}\label{gif}
\end{center}
\end{figure*} 
(e) {\it Stripe phase}: the filled sites form diagonal stripes (Fig.2c second 
column). This phase is observed at higher values of $U$ (above $U \sim 5$) 
at $n_f=1/2$. It is found to be stable at finite $t^\prime$ ranging 
from $-1$ to $0.8$ beyond which there is a phase segregation. Comparing 
our results with earlier results at $t^\prime=0$ limit~\cite{farkov_pss}, we 
find that there can be a phase segregation now at a lower value of $U$ 
(for example, $U\simeq 1$) as the parameter $t^\prime$ is tuned. 
The $t^\prime$ term would, therefore, play an important role in driving 
phase segregation even in the weakly correlated systems. This could be 
understood by looking at the position
of the localized $f$-electron level ($E_f$) with respect to  the $d$-electron 
band. At large $U$, the $d$-band is strongly renormalized and the modulation 
of the $d$-band due to the $t^\prime$ term becomes insignificant. It is also
noticed that $t^{\prime}=1$ is a special point as at this point hopping 
takes place only between pairs of sites having identical $f$-electron occupation
(near-neighbour configurations $\{0,0\}$ and $\{1,1\}$, though, occupying different  
regions of the Brillouin zone). This enhances the formation of inhomogeneous 
phases where a pair of $f$-electrons (or a pair of empty sites) tend to 
occupy neighboring places on the lattice as in Fig.2 column 3. If the 
coherence of the $f$-electrons is restored, this would lead to 
the formation of interband excitons involving near-neighbor sites (at finite 
$f-d$ mixing). In the context of Hubbard model, where $d\, 
(f)$-electrons represent 
up (down) spins, the correlated hopping would, therefore, tend to enhance 
the extended s-wave superconducting order parameter (OP) fluctuations.   


Shown in Fig.2 are some of the ground state patterns mentioned above and 
their dependence on the correlated hopping parameter $t^{\prime}$ as well as the 
Coulomb correlation strength $U$. Fig.2(a) (top panel) reveals the variety of
ground state patterns at $n_{f}=1/4$ starting from regular structures  
to phase segregated regions on changing the value of $t^\prime$ 
from $-1$ to $+1$ at $U=1$. This is also observed at $n_{f}=1/3$ where on 
varying either $U$ or $t^\prime$ one could go from a quasi-regular pattern to 
phase segregated regions. At $n_{f}=\frac{1}{2}$ (panel (c)) 
we have shown three patterns corresponding to three different sets of values 
for $U$ and $t^\prime$ 
(i.e., bounded, stripe and phase segregated). On changing the value of $U$ 
from 1 to 5, keeping $t^\prime$ fixed at $-1$, a transition from a bounded 
phase to a stripe phase occurs whereas on changing the value of $t^\prime$ 
from -1 to +1 at a fixed $U=1$, we arrive at a phase segregated state. And 
finally in panel (d) we 
present the phases observed at $n_{f}=2/3$, ranging from bounded, hexagonal 
to phase segregated regions at three different $t^\prime$. It is clear from
the figure that a non-zero $t^\prime$ facilitates phase segregation in the 
ground state as it favors simultaneous occupation of $f$-electrons 
(or simultaneous occupation of vacancy, depending on the value of $t^\prime$) 
in the neighboring sites. Phase segregations have been the key to the physics 
of many correlated systems hotly pursued in the last decade~\cite{dagbook} and 
this correlated hopping term seems to open up another route to the  
phase segregation scenario in certain correlated systems.  
A ground state phase diagram in the $t^{\prime}-U$ plane could be drawn 
now, based on the structures obtained in Fig.2. We show this at $n_{f}=
\frac{2}{3} $ in Fig.3. One can identify the three different phases, 
namely, {\it bounded}, {\it hexagonal} and {\it segregated}.  
The reasonably large region of phase segregations at the  
upper part of the phase diagram owes its origin to $t^\prime$ primarily as 
discussed earlier. At large $U$, of course, the hexagonal phase dominates.  
At this filling, no other phases were found in the range of parameters 
studied. 
\begin{figure*}[ht]
\begin{center}
\includegraphics[width=14cm]{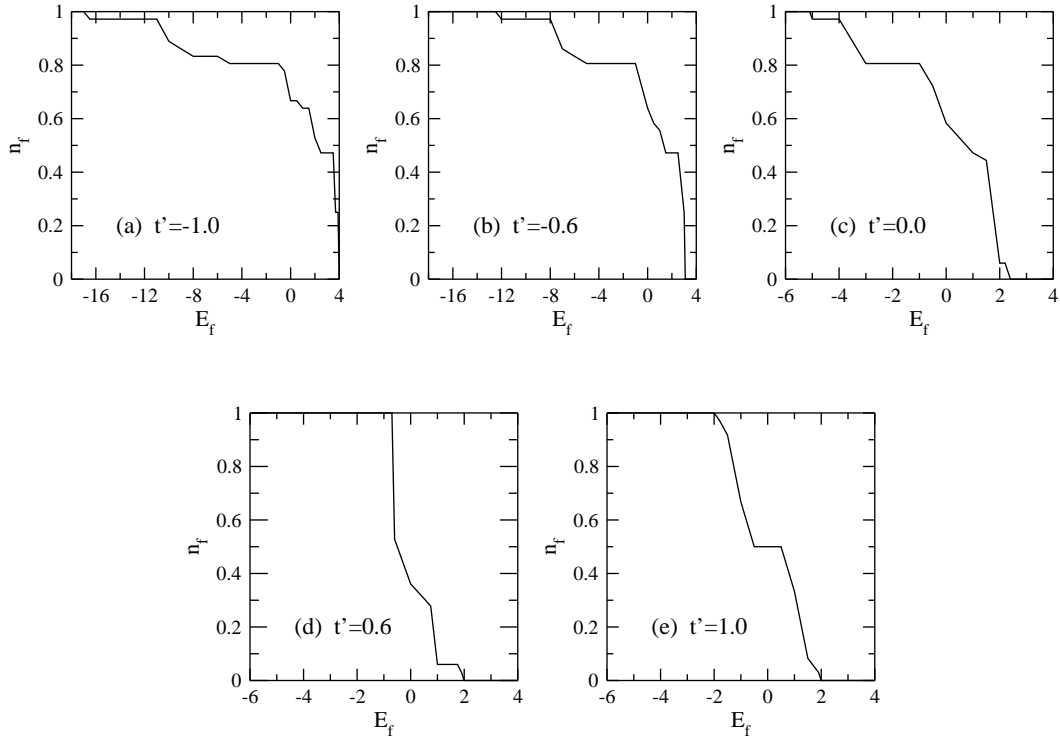}
\caption{$n_{f}-E_{f}$ phase diagrams for $U=1$ and different values of 
$t^{\prime}=-1, -0.6, 0, 0.6$ and $1$ are shown in (a).....(e)
respectively.}
\label{fig6}
\end{center}
\end{figure*}
 
In order to identify if there is a metal to insulator transition in the above 
phase diagrams, we look at the $d$-electron spectrum (Fig.4) and observe the 
gap~\cite{gap_defn} at the Fermi level at various values of $U$ and $t^\prime$
in the corresponding density of states (DOS). In Fig.4(a) is shown the DOS for 
three distinct phases observed 
at $n_{f}=2/3$ keeping $U$ fixed while changing $t^\prime$. Fig.4(b) shows the
DOS for different values of $U, \, t^\prime$ in two different phases 
at $n_{f}=\frac{1}{4}$. We find that the
hexagonal phase at $U=5$ is insulating with a large gap, the bounded phase at 
low $U$ is also insulating but with a small gap while the phase segregated 
regions appear to have a very small (or vanishing) gap. 
Although the charge excitation gap continues to increase with $U$ in
the insulating phases, a `spin' excitation involving exchange between vacant
and occupied $f$-electron sites has a lower energy scale $\sim \frac{1}{U}$
at large $U$.  In this context, we note
the evolution of the spectrum and the corresponding DOS
in Fig.4(b) at a lower $n_{f}=1/4$. It is clear that the DOS in the regular phases
($U=1,t^{\prime}=-1$ and $U=5,t^{\prime}=-1$) have 
two-peak like structures above and below Fermi energy $E_F$ (in the
phases with a clear gap at the Fermi level) and the gap
at the Fermi level increases with $U$. The band spreads
out towards higher energy as expected. But the segregated phase 
($U=1,t^{\prime}=1$) here is observed to have an almost gapless spectrum.
Whether a very tiny gap still exists is not clear from this finite-size
calculations, but within our numerical accuracy it appears to have closed.  
Such observations are reported earlier, albeit on a square lattice away 
from half-filling~\cite{lach}. It is possible that due to the segregation 
of $f$-electrons in one part of the 
lattice, $d$-electrons find a percolative path in the rest of the lattice and move 
freely from one end of the lattice to other without appreciable effects of 
correlation in the segregated phase.

It is clear that the degeneracies as well as the separation between the
energy levels increase as $U$ rises and lead to incompressible states 
at specific values of the filling (or, equivalently, the chemical potential) 
in the thermodynamic limit. The FKM is known to show first order transitions and 
consequent phase separations in the bipartite lattices and it appears similar 
physics holds good in large part of the phase diagram for a triangular 
lattice as well. In order to glean a physical picture of the local electron 
occupancies in different
regimes, we have drawn the density of the $d$-electrons on the real
lattice (Fig.5) where the radius of the circles are proportional to $d$-electron
density at that site (a color coding is also used). A change of $t^\prime$ from
$1$ to $-1$ leads to an increase in $d$-electron density on the unoccupied
sites (sites without an $f$-electron) as discussed above. For $t^{\prime}=1$ 
there is a reasonable weight of the $d$-electrons in the occupied sites at
smaller $U$. As $U$ increases the wave functions spread out from occupied
sites. Thus the competition between localization and itinerancy is clearly
visible from these figures.

The FKM is known to have transitions~\cite{free} involving states 
with different
$n_f$ values. We have looked at these transitions as a function of $N_f$ 
(keeping $N_{f}+N_{d}=N$) depending on the relative position of 
the $f$-electron level 
($E_f$) and the Fermi level of the $d$-electrons. If $E_f$ lies above 
the Fermi energy ($E_F$), then $N_{d}=N$ and the $f$-states are empty. In the 
opposite limit, we are in a classical mixed-valent regime. For a given set 
of $t^{\prime},\, U$ and $E_{f}$, one can easily find out the values 
of $N_{f}$ for which the ground state 
has minimum energy. We have plotted $n_{f}$ in Fig.6 as a 
function of $E_{f}$ for a set of values of $t^{\prime}$ and $U$.

Figs.6(a)...6(e) show $n_f-E_f$ phase diagrams for $U=1$ and correlated 
hopping $t^{\prime}=-1.0, -0.6, 0, 0.6$, and $1.0$ respectively. We observe 
that as $t^{\prime}$ increases from $-1.0$ to $+0.6$, the valence 
transition (i.e. $n_f$-transition) (i) occurs at smaller values of $E_f$ and 
(ii) becomes sharper. However, as shown in Fig.6(e), the $n_{f}$-transition 
becomes smoother again for $t^{\prime}$=$1.0$. We observe that the 
transition width 
(the range of $E_{f}$ over which $n_{f}$ goes from $1.0$ to $0.0$), decreases 
as $t^{\prime}$ increases from $-1.0$ to $0.6$. This could be explained from 
the fact that at $t^{\prime}=-1.0$, the position of $E_{f}$, where 
$n_{f}$ starts decreasing (from its maximum value 1.0) is located far below the 
centre of the $d$-band. As $t^{\prime}$ increases towards zero the position of 
$E_{f}$, at which $n_{f}$ starts decreasing, moves towards the centre 
of $d$-band. For $t^{\prime} >0.6$, the corresponding position of $E_{f}$ 
again moves below the centre of $d$-band. So the effective width of $d$-band 
is maximum for $t^{\prime}=-1.0$, is minimum around $t^{\prime}=0.5$ and 
increases again for $t^{\prime} >0.5$.
In Fig.7(a), 7(b) and 7(c) we show the same for $U=1, 5, 10$ with 
$t^{\prime}=-0.6,\, 0$ and $0.4$. We observe here that the transition 
width decreases 
as the value of $U$ increases. The value of $E_{f}$ where major change in $n_{f}$ 
occurs hardly depends on the value of $U$. For all values of $U$, the 
$n_{f}$-transition is relatively smooth for $t^{\prime}$ around $-0.6$ and 
this transition becomes steep for $t^{\prime}$ around $0.4$. These features are
easily understood from an analysis of the spectrum of $d$-electrons.  

As an interesting aside, we note that although the extended FKM studied 
here has $U>0$, the corresponding ground state phase diagram for $U<0$ is 
obtainable from the following symmetry: if $\{\omega_i\}$ is the 
ground state for a
particular $\mu, E_f$ and $U$, then  $\{\bar{\omega}_i\}$ is the ground 
state for  $\mu, -E_f$, $-t^{\prime}$ and $-U$, 
where $\bar{\omega}_{i}=1-\omega_{i}$. The 
negative $U$ FKM was introduced by Lieb~\cite{lieb} as a model for the study 
of the formation of crystalline solids. In that context, the heavy $f$-electrons 
can be thought to represent the ions in a solid, and their ordered structure 
implies a crystalline arrangement of the lattice.    
\begin{figure*}[ht]
\begin{center}
\includegraphics[width=15cm]{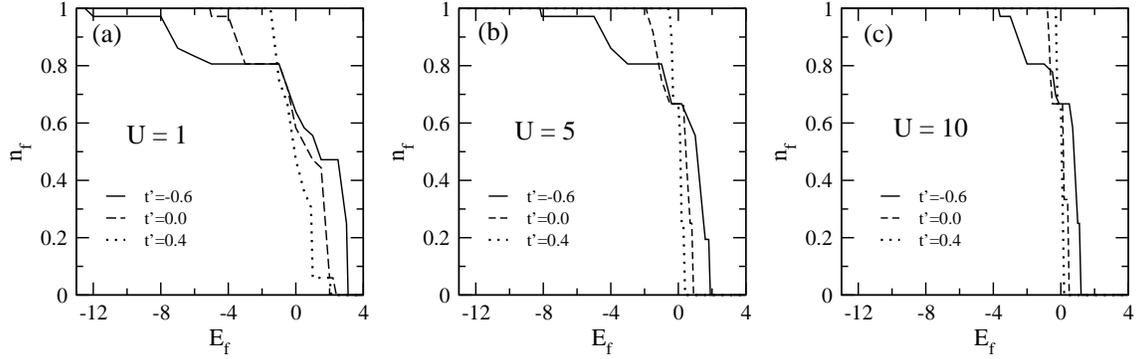}
\caption{$n_{f}-E_{f}$ phase diagram for different values of
 correlated hopping $t^{\prime}=-0.6, 0$ and $0.4$,\, 
 at (a) $U=1$ , (b) $U=5$  and (c) $U=10$.}\label{gif}
\end{center}
\end{figure*}

It is in order now to discuss the numerical results obtained above 
in the context of the systems where such models are expected to be 
useful under simplifying assumptions, in certain regions of their phase diagrams.    
Fig.1 gives a good account of the complexity of phases obtainable from an FKM on 
a triangular lattice. The model produces charge ordered states of varying 
periodicities and phase segregated states as filling, $U$ and $t^\prime$ are 
changed, also observed in dichalcogenides, cobaltates and GdI$_2$. 
The hexagonal structure close to $n_{f}=2/3$ is reminiscent~\cite{uy_tobe}
of the superlattice structure in Na$_x$CoO$_2$ (at $x=2/3$ in 
Fig.3(c) of Ref. 4). The CDW in 2H-TaS$_2$ is incommensurate~\cite{cava}, 
while it is commensurate in TiSe$_2$~\cite{qian2}, though the superlattice 
periodicity is dictated by special relations between the nearly flat valence band 
and the conduction band~\cite{aebi,qian2}, rather than the geometry of the
underlying lattice, since the CDW in TiSe$_2$ is likely to be excitonic in 
origin~\cite{aebi}. Stripes in correlated systems tend to appear at special 
fillings~\cite{dagbook} due to commensurability effects in bipartite lattices, 
while we observe its formation here (e.g., at $n_{f}=\frac{1}{2}$) even if the 
underlying lattice is non-bipartite. Stripes have not yet been seen experimentally 
in the systems 
we discussed, but appear to be a distinct possibility if carefully investigated. 
The phases revealed here clearly shows the intricate 
nature of the competing states alluded to earlier - correlation and frustration 
working together on a triangular lattice in unraveling such a rich phase diagram. 
This underlines the need for a thorough experimental investigation of these 
systems to delineate both the ordered and inhomogeneous phases as filling and 
parameters change. 

As discussed earlier, a common feature among systems like dichalcogenides
~\cite{qian2,cava} and Na$_x$CoO$_2$\, ~\cite{qian} is the competition between 
a charge density order and superconductivity, where slight changes in symmetry 
or a small perturbation may lead to a transition from one to the other. 
Off-diagonal long range order (ODLRO) is absent in the FK model unless 
of course the $f$-electrons become dispersive. In the limit of infinitely 
heavy down-spin electrons ($t_{ij,\downarrow}=0$), the Hubbard model 
reduces to the (symmetric) FKM. With any symmetry-breaking perturbation 
(pressure~\cite{aebi}, doping/hydration~\cite{qian2,qian,cava} etc.), 
as $t_{ij,\downarrow}$ shifts away from zero, the charge order 
or phase segregation may disappear~\cite{free-lieb} and superconductivity 
could appear: $s$-wave for the attractive-$U$ model and extended $s$-wave 
in the repulsive-$U$ case, due explicitly to the $t^{\prime}$ term~\cite{hirsch2}. 
The nature of superconductivity in both these systems should, therefore, shed 
considerable light on the underlying microscopic mechanism operative here. 
Although there is a suggestion of high angular momentum superconducting OP in 
doped dichalcogenides~\cite{castro}, the symmetries of OP in hydrated cobaltate and 
doped dichalcogenides are not resolved yet and should be probed thoroughly.  
 
In GdI$_2$ a phase segregation appears to preempt the formation of a long range 
order. While phase segregation has been shown to appear in FKM on bipartite 
lattices~\cite{free-lieb, kennedy2}, it requires a fairly strong $U$. GdI$_2$ 
falls in the intermediate coupling range and the phase segregation seen in 
GdI$_2$ at low temperatures even for a low $U$ could be facilitated by the 
presence of $t^{\prime}$ as shown above. The correlated hopping term helps in  
tuning the balance between states close in energy, and therefore, plays  
a significant role in non-bipartite lattices where nearly degenerate states 
proliferate at low temperatures. We note that a quantitative modeling in detail,
including first principles band 
structure and Fermi surface topologies, is beyond the present model. Such 
a study is indeed very useful and left for the future. We believe our present 
study already sheds light on certain important aspects of the ground state 
phenomenology and motivates new experiments. 

In conclusion, we have studied the Falicov-Kimball model on a 
non-bipartite lattice and found several ordered ground states. 
Extending the model to include a correlated hopping term leads to new 
effects: in particular, it strongly facilitates phase segregation
in the ground states even in the weak correlation limit. We also 
identify several valence transitions as $t^\prime$ and $U$ are 
changed. These observations are quite relevant for the ground state 
order and phase segregation observed recently in correlated systems with 
triangular lattices such as $GdI_2$, dichalcogenides and cobaltates.
\vspace{0.2cm}

\noindent{\bf Acknowledgement}
TM acknowledges support from the Max Planck Institute for Physics of Complex
Systems, Dresden, Germany. UKY acknowledges CSIR, India for research 
fellowship. AT acknowledges useful discussions with M. Laad. 

\end{document}